\documentclass[10pt,twocolumn,twoside]{IEEEtran}
\usepackage{amsfonts}
\usepackage{amsmath}
\usepackage{hyperref}
\usepackage{amsmath,amsfonts,amssymb,bm}
\usepackage{tikz,pgf}

\usepackage{pstricks}
\newtheorem{definition}{Definition}
\newtheorem{problem}{Problem}
\newtheorem{proof}{Proof}

\newcommand{\be}{\begin{eqnarray}}
\newcommand{\ee}{\end{eqnarray}}
\newcommand{\bi}{\begin{itemize}}
\newcommand{\ei}{\end{itemize}}
\newcommand{\ba}{\begin{array}}
\newcommand{\ea}{\end{array}}
\newcommand{\bbs}{\begin{slide}}
\newcommand{\es}{\end{slide}}

\usepackage{graphicx}
\usepackage{subcaption}

\ifCLASSINFOpdf
\else
\fi
\begin{document}

%
\title{Consensus analysis of a two-step communication opinion dynamics model with group pressure and self-confidence}
%
%
%

\author{Wenjuan Wang, Zhongmei Wang, and Xinmin Song
        \thanks{This work was supported by the National Natural Science Foundation of China under Grants 61803239, 61873152,61821004, 62250056, and the Natural Science Foundation of Shandong Province (ZR2021ZD14, ZR2021JQ24), and Highlevel Talent Team Project of Qingdao West Coast New Area (RCTD-JC-2019-05).

        W. Wang is with the Business School, Shandong Normal University,  Jinan, Shandong, P.R.China 250000.
        Z. Wang is with the Institute of System and Control Science, Shandong Normal University,  Jinan, Shandong, P.R.China 250000.
        X. Song is with the School of Information Science and Engineering, Shandong Normal University,  Jinan, Shandong, P.R.China 250000.

({\tt\small  e-mail: wangwenjuan2244@163.com; wangzhongmei211@163.com; xinminsong@sina.com}).}}
\maketitle

\begin{abstract}
This paper considers the consensus problem of a novel opinion dynamics model with group pressure and self-confidence. Different with the most existing paper,  the influence of friends of  friends in a social network is taken into account, which is modeled to be  two-step communication. Based on this consideration,  the neighbors of agents are classified into direct neighbors and indirect neighbors. Accordingly, the communication between agents and their neighbors is classified into one-step communication and  two-step communication. By applying matrix analytic theory and graph theory, it is shown that the opinion consensus can be achieved.  Moreover, the exactly consensus value of the opinion is obtained for three cases of the group pressure. Finally, simulation examples are provided to demonstrate the validity of the conclusions drawn in the paper.
\end{abstract}


%
\IEEEpeerreviewmaketitle

\section{Introduction}
\IEEEPARstart{R}{ecently},
 opinion dynamics has rapidly developed and attracted a lot of attention  from control theory, physics, social psychology  and so on \cite{TIAN2018213,wulixue,dandekar2013biased,10049708,li2019clustering}. In a social network, individuals generate or update their own opinions by communicating with others.
As to a particular topic, the opinions of different individuals  propagate on the social network and influence each other, based on which a collective opinion  on this topic is formed.

Early work on social networks focused on quantitative methods (later called sociometry) to describe social relations. In 1934, Moreno \cite{citation-key} first introduced a graphical tool to visualize the underlying structure of groups and the positions of each individual,  which was later termed networks. Subsequently, the term ``social network" was proposed. In 1948, the pioneering work \cite{wiener2019cybernetics} in cybernetics laid the foundation for the emergence of sociocybernetics \cite{geyer1995challenge}. The combination  of sociology and cybernetics  has shifted the focus of social network research towards the study of dynamic systems perspective \cite{zheng1992current}. This has given rise to a new research direction known as opinion dynamics. In recent years, the development of multi-agent systems has provided  rich  mathematical models for opinion dynamics
\cite{jin2023distributed,
qin2021multiagent,
multi2021,valcher2017consensus,liu2023event}.

Several classical opinion dynamic models are formulated in \cite{degroot1974reaching,friedkin1990social,deffuant2000mixing,hegselmann2002opinion}. The DeGroot model was proposed by DeGroot in 1974 \cite{degroot1974reaching}. In 1990, self-confidence was considered into the Degroot model, which is known as Friedkin-Johnsen (FJ) model \cite{friedkin1990social}. Different from \cite{friedkin1990social}, bounded confidence models were proposed in \cite{deffuant2000mixing} and \cite{hegselmann2002opinion}, which are known as Deffuant-Weisbuch (DW) model and Hegselmann-Krause (HK) model, respectively. In these two models, individual opinions can only be influenced by  neighbors whose opinion difference is within the confidence threshold.

The classical opinion  models mentioned above provide fundamental support for the investigation of opinion dynamics. As one of the basic problems in the study of  opinion dynamics, opinion consensus has received considerable attention in recent years \cite{wanglong,bernardo2024bounded,bipartite2021,expressed}.
To explain the complex phenomenon and predict opinion fluctuation,  \cite{xing2021novel}  proposed a novel model by unifying  the Degroot model and FJ model, in which necessary and sufficient conditions are presented for the opinion consensus of the proposed model.
\cite{su2017noise} considered the consensus behavior of the HK model in noisy environments. It is demonstrated that when noise strength is below the critical value,  opinions in the noisy HK model  tend to reach  consensus in  finite time. \cite{he2021discrete} considered two discrete-time signed bounded confidence models and provided  necessary and sufficient conditions for the bipartite consensus.

In the real world, an individual's decision is inevitably affected by conformity \cite{anderson2019recent,ye2019influence,javarone2014social}.  To characterize the conformity, \cite{Cheng2019}  introduced group pressure into  opinion dynamics based on the HK model and theoretically proved that all individuals can reach a consensus in  finite time. Inspired by the former's work, \cite{Zhang2024}  considered  a modified HK model in which the weights of individuals influenced  by others are not always the same.

With the rapid development of the Internet, more and more social platforms (e.g., Twitter, Facebook and Weibo) arise, which provide people a convenient  communication channel.
Through these platforms,  high-order interactions can be made \cite{qian2011adaptive}. In detail, people can not only communicate with their friends, but also interact with friends of friends \cite{connected}. It has been found that the relationships between individuals and those two degrees of separation away have a strong correlation.
To describe this kind of high-order interaction,
\cite{atwostep2020}  proposed a two-step communication opinion dynamics model  based on the DeGroot model, in which the influence of friends' friends  is taken into consideration.  Furthermore, self-confidence is introduced in \cite{atwostep2020}  to measure the individual adherence to the initial opinion.
\cite{DONG2022218} suggests that two individuals who are strangers can establish a connection and become friends through common friends acting as bridge nodes.
In fact, both ``high-order model" proposed in \cite{qian2011adaptive}, the method ``friends of friends" introduced in \cite{connected} and bridge nodes (common neighbors) in \cite{DONG2022218}  are all two-step communication.

In many scenarios, self-confidence, group pressure and  two-step communication affect the evolution of opinions simultaneously.  However, to our knowledge,  there are no results concerning these three factors simultaneously in the existing work.
In this paper, we consider the consensus analysis of a two-step communication opinion dynamics model with group
pressure and self-confidence. The opinion dynamics  model in this paper is formulated based on the DeGroot model.
The contributions are as follows.

(1)A two-step communication opinion dynamics model is proposed based on the DeGroot model, in which the self-confidence and group pressure are considered simultaneously. Different from \cite{atwostep2020}, the influence of group pressure on opinion dynamics is introduced in this novel model. In addition,
a new weight matrix calculation method is proposed, where individuals assign
the weight to their neighbors based on node degrees.

(2)By applying stochastic matrix theory and graph theory, it is proved that the opinions  can
reach a consensus. Furthermore, the consensus value under different group pressure level is given exactly.
Instead of introducing group pressure into the HK model in \cite{Cheng2019,Zhang2024}, the gap was filled by considering group pressure based on the DeGroot model.

The rest of this paper is organized as follows. Section II  introduces some basic concepts about graph theory  and the DeGroot model. Section III  formulates a novel opinion dynamics model to be studied in this paper.
The consensus results of this model are presented in Section IV.
In Section V, simulation examples are presented to verify the effectiveness of the theoretical results. Finally, Section VI  provides conclusions of this paper.

\textit{Notations:}
Standard notations are used throughout this paper: the symbol $W^{t}$ represents the matrix $W$ raised to the power of $t$. $\mathbf{1}$ is the  $n \times 1$ vector with all entries equal to $1$. A fixed matrix refers to a matrix whose elements remain constant and do not change over time. For a vector $\alpha$, $\alpha^{T}$ denotes the transpose of $\alpha$. $\emptyset$ denotes the empty set.

%

%
%
%
\section{Preliminaries}

\subsection{Graph theory}
Let  $G(V, E,A)$ be a directed graph in which $V=\{1,2,...,n\}$ denotes the set of finite nodes,
$E\subseteq V \times V$ denotes
the set of edges.
An edge $e_{ij}\in E$  represents a directional relationship from node $i$ to node $j$, indicating that node $i$ trusts node $j$ and obtains information from it, but not necessarily vice versa.
The adjacency matrix of $G(V,E,A)$ is denoted $A= (a_{ij})_{n\times n}$, where $a_{ij}=1$ if and
only if $e_{ij}\in E$, otherwise $a_{ij}=0$. The indegree is the sum of incoming edges to $i$, denoted as $deg^{+}_{i}=\sum_{k=1}^{n}a_{ki}$, while the outdegree is the sum of outgoing edges from $i$, denoted as $deg^{-}_{i}=\sum_{k=1}^{n}a_{ik}$.
In this paper, it is assumed that  each agent must trust at least one agent except himself, i.e., $deg^{-}_{i}\geq1$. Let $P= (p_{ij})_{n\times n}$ be the accessibility matrix of $G(V, E,A)$, if there exists a path from node $i$ to node $j$, then $p_{ij}=1$, otherwise $p_{ij}=0$. Specially,  $p_{ii}=1$ indicates the presence of a self-loop where node $i$ can reach itself.
\subsection{ Introduction of DeGroot model}
Since the opinion dynamics model to be proposed in this paper is formulated based on the DeGroot  model, we firstly review the DeGroot model  as follows \cite{degroot1974reaching}.

Consider a system of $n$ agents forming opinions in a network.
The opinion value of agent $i$ at time $t$ is denoted as  $x_{i}(t)$, which satisfies that
$0 \le x_{i}(t) \le 1$.
In  the DeGroot model, the opinion dynamics of agent $i$ is described as follows.

\be\label{equ1}
  x_{i} (t+1) = w_{i1} x_{1} (t)+w_{i2} x_{2} (t)+...+w_{in} x_{n} (t),
\ee
where $w_{ij}$ $ \in $ [0,1] represents the  weight assigned by agent $i$ to agent $j$. For each agent,  it holds that $\sum_{j=1}^{n} w_{ij}=1,  (i=1,2,...,n).$
The DeGroot model  assumes that the social network is time-invariant, in which each agent opinion is determined by a weighted average of their own opinion and that of their neighbors. Let $x(t)$ = $(x_{1}(t),x_{2}(t), ... ,x_{n}(t))^{T}\in \mathbb{R}^n$ denote the opinion profile of all agents at time $t$. Therefore, the
system (\ref{equ1}) is expressed in matrix form:
\be\label{equ2}
  x(t+1)=Wx(t),t=0,1,2...
\ee
where $W= (w_{ij})_{n \times n}$ is a stochastic and fixed matrix. It is noted that the DeGroot model is a linear model. As  opinions evolves over time, the evolution of opinions is as follows:
\be\label{equ3}
  x(t)=Wx(t-1)=...=W^{t}x(0),t=0,1,2...
\ee

\section{Problem formulation}

In this section, some  definitions related to the study are presented firstly.

\begin{definition}
\textnormal{\cite{Cheng2019} Group pressure denotes the influence exerted by a social group on its individuals to conform to the group opinions, behavior or beliefs.}
\end{definition}
\newtheorem{definition2}{Definition2}
\begin{definition}
\textnormal{For an agent, its neighbors are classified to be the following two groups. \\
(1) For agent $i$ and agent $j$, define  agent $j$ is a direct neighbor of agent $i$  if $a_{ij}=1$.
The set formed by all direct neighbors of
$i$ is denoted as $\mathcal{N}_{1}(i)$, where $\mathcal{N}_{1}(i)=\{j\in V\mid a_{ij}=1\}$.\\
(2) For agent $i$ , agent $j$ and agent $s$, define agent $s$ is an indirect neighbor of agent $i$ if $j\in \mathcal{N}_{1}(i)$ and $s\in \mathcal{N}_{1}(j)$, but $s\notin \mathcal{N}_{1}(i)$. The set formed by all indirect neighbors of  $i$  is denoted as $\mathcal{N}_{2}(i)$, where $\mathcal{N}_{2}(i)=
\{s\in V\mid a_{ij}=1, a_{js}=1, a_{is}=0\}$.}
\end{definition}
\begin{definition}
\textnormal{\cite{atwostep2020}
The communication of agents is classified to be the following two cases.\\
(1) One-step communication refers that  agents trust only their direct neighbors. In detail, the opinion of an agent is influenced by only its  direct neighbors.\\
(2) Two-step communication refers that agents  trust not only  their direct neighbors but also indirect neighbors. In detail, the opinion of an agent is influenced by both direct neighbors and indirect neighbors.}
\end{definition}

Combining self-confidence, group pressure and two-step communication defined as above, a novel opinion dynamics model is introduced as follows.
\begin{equation}\label{equ4}
  x_{i}(t+1)=f_{i}x_{avg}(t)+(1-f_{i})(\alpha _{i} x_{i} (t)+ {\textstyle \sum_{j=1}^{n}} w_{ij}^{'}x_{j }(t)),
\end{equation}
where $i,j\in V, i\neq j$; $f_{i}\in[0,1]$ denotes the group pressure of agent $i$, while $1-f_{i}$ is the resilience  to this pressure of agent $i$. $x_{avg}(t)=\frac{1}{n}  {\textstyle \sum_{k\in V}} x_{k}(t)$ is the average opinion of the group at time $t$, representing the public opinion. $\alpha_{i}\in(0,1)$ represents  self-confidence of  agent $i$.
$w_{ij}^{'}$ represents the weight assigned by agent $i$ to agent $j$ in the proposed  model, specifically,
\begin{equation}\label{equ5}
w_{ij}^{'}=
\left\{\begin{matrix} (con_{i}\times INF_{j}-w_{ij})\times r+w_{ij},  &j\in \mathcal{N}_{1}(i);& \\  con_{i}\times INF_{j} \times r \times \frac{deg^{+}_{j}}{\sum\limits _{k\in\,\mathcal{N}_{2}(i)}deg^{+}_{k}},  & j\in \mathcal{N}_{2}(i); & \\\alpha_{i},& i=j;&\\0,  & \mbox{otherwise}. &\end{matrix}\right.
\end{equation}
where $r\in (0,1]$ is applied to characterize the extent to which  an agent desires to communicate  with its indirect neighbors. $INF_{i}\triangleq\frac{\alpha_{i}+\beta_{i}+\gamma_{i}}{3}$ represents the influence index of agent $i$ in the group; $\beta_{i}\triangleq\frac{deg_i^+}{n-1}$ denotes the in-degree centrality of agent $i$ and $\gamma_{i}\triangleq\frac{deg_i^-}{n-1}$ denotes the out-degree centrality of agent $i$,  which is defined similar to \cite{atwostep2020}. In a directed graph, $\beta_{i}$ and $\gamma_{i}$ respectively characterize the degree to which an agent influences others and is influenced by others.
\begin{align}\label{equ6}
con_{i}&=\frac{1-\alpha_{i}}{\sum INF}; \nonumber \\
\sum INF&=\sum _{r\in\,\mathcal{N}_{1}(i) }INF_{i}+\sum _{s\in\,\mathcal{N}_{2}(i) }INF_{s}\frac{deg_{s}^+}{\sum\limits_{k\in\,\mathcal{N}_{2}(i) } deg_{k}^+};\nonumber \\
w_{ij}&=\frac{INF_{j}}{\sum\limits _{k\in\,\mathcal{N}_{1}(i)}INF_{k}} (1-\alpha _{i})a_{ij}.
\end{align}

\begin{figure}[ht]
\centering
\includegraphics[width=0.25\textwidth]{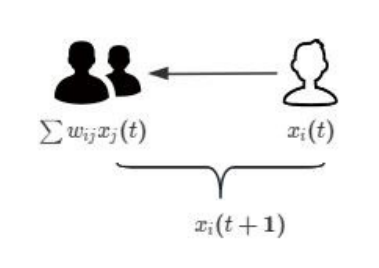}
\caption{Block diagram of opinion dynamics in traditional model.}
\label{fig:1}
\end{figure}

\begin{figure}[ht]
\centering
\includegraphics[width=0.4\textwidth]{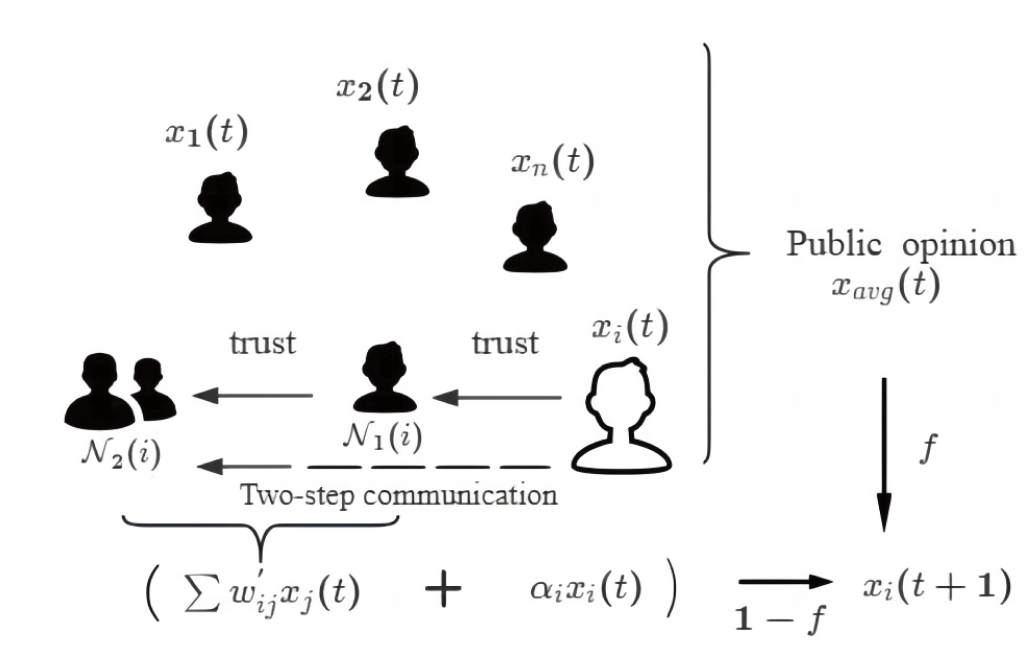}
\caption{Block diagram of opinion dynamics in the system (\ref{equ4}).}
\label{fig:2}
\end{figure}

Fig. \ref{fig:1} presents the opinion evolution of the DeGroot model.
Fig. \ref{fig:2}  presents the opinion evolution of the model (\ref{equ4}). It is worth mentioning that the proposed opinion dynamics model extends the trust scope of agents to their indirect neighbors, thereby  allowing  traditional first-order  interaction into higher-order structures. The realism  of this model are much closer to actual social networks.

Opinion consensus is to be considered in this paper, the definition of consensus is given as follows.
\newtheorem{definition5(Convergence)}{Definition5(Convergence)}
\begin{definition}
\textnormal{For any initial condition $x(0)\in\mathbb{R}^n$, there  exists a constant vector $x^{*}\in\mathbb{R}^n$,
such that $\lim_{t \to \infty}x(t)=x^{*}$, we say the  system (4) converges. The convergent system achieves consensus if for any opinion vector $x(0)\in\mathbb{R}^n$, $\lim_{t \to \infty}x(t)= c\mathbf{1}$, where $c\in \mathbb{R} $ is a constant and $\mathbf{1}$ is  the $n \times 1$ vector with all entries equal to $1$.}
\end{definition}
\begin{problem}
 \textnormal{In this paper, the objective is to study the consensusability of  the opinion dynamics model (\ref{equ4}).  }
\end{problem}
\section{Consensus analysis}

In this section, we will analyze the consensusability of the system (\ref{equ4}). Without loss of generality, it is assumed that $f_{1} = f_{2} =...=f_{n} =f$ in this paper. The following Lemma is necessary.
\newtheorem{Lemma}{Lemma}
\begin{Lemma}
\label{le1}
\textnormal{\cite{dong2017managing}} In the DeGroot model, all agents will reach a consensus if and only if there exists a  $t^{*}\in \mathbb{N}^{+}$ such that the matrix power $B^{t^*}$ has at least one column with all elements strictly positive.
\end{Lemma}
\newtheorem{Theorem}{Theorem}
\begin{Theorem}
\label{th1}
For the system (\ref{equ4}), the following consensus conclusions are obtained.\\
(1) When $f=0$, $\lim_{t\to\infty} x_{i}(t)=\sum\limits_{i\in V_{c} }\frac{\pi_{i} }{\sum_{i\in V_{c}}\pi_{i} } x_{i}(0)$, where $V_{c}=\{i\in V\mid p_{ji}=1, j=1,2,...,n\}$, and $p_{ji}$ is the element of the accessibility matrix $P$.\\
(2) When $f=1$,  consensus  is achieved at  time $t=1$. Thus $\lim\limits_{t\to\infty} x_{i}(t)=\frac{1}{n}  {\textstyle \sum_{j=1}^{n}} x_{j}(0)$.\\
(3) When $f\in (0,1)$, $\lim\limits_{t\to\infty} x_{i}(t)={\textstyle \sum_{i=1}^{n}} \lambda_{i}x_{i}(0)$, where the non-negative vector $\lambda\in \mathbb{R}^{n}$ satisfies $\lambda\, B=\lambda$ and  ${\textstyle \sum_{i=1}^{n}} \lambda_{i}=1$. Here \begin{align}
B=\begin{pmatrix}(1-f)\alpha_{1}+\frac{f}{n}   & \dots  & (1-f)w_{1n}^{'}+\frac{f}{n} \\\vdots & \ddots   &\vdots  \\(1-f)w_{n1}^{'}+\frac{f}{n}   & \cdots  &(1-f)\alpha_{n}+\frac{f}{n}\end{pmatrix}.\nonumber
\end{align}
\end{Theorem}
\newtheorem{Proof}{Proof}
\begin{proof}
Case (1): The opinions of agents are not influenced by group pressure, i.e. $f=0$. Similar to the proof in \cite{atwostep2020},  the sufficient condition for the model (\ref{equ4}) to achieve consensus is as follows.

If there exists at least one
celebrity $i$  in the social network such that for any agent $j$, it holds that $p_{ji}=1$, then the model (\ref{equ4}) can achieve consensus  and the consensus value is $c=\sum_{i\in V_{c} }\frac{\pi_{i} }{\sum_{i\in V_{c}}\pi_{i} } x_{i}(0)$, where $V_{c}=\{i\in V\mid p_{ji}=1, j=1,2,...,n\}$, and $p_{ji}$ is the element of the accessibility matrix $P$.

Case (2): The opinions of agents are  influenced to the maximum extent by group pressure, i.e. $f=1$.  Eq. (\ref{equ4}) is rewritten as:
\begin{equation}\label{7}
  x_{i}(t+1)= x_{avg}(t)=\frac{1}{n}  {\textstyle \sum_{j=1}^{n}} x_{j}(t).
\end{equation}
For each agent, it is noted that\begin{equation}\label{8}
 x_{1}(1)=x_{2}(1)=...=x_{n}(1)=\frac{1}{n}  {\textstyle \sum_{j=1}^{n}} x_{j}(0).
\end{equation}
The model (\ref{equ4}) achieves consensus at time  $t=1$, where each agent updates their opinion to the average of all agents' original opinions. Consequently, the consensus opinion of the group is $c=\frac{1}{n}  {\textstyle \sum_{j=1}^{n}} x_{j}(0)$.

Case (3): When $f\in (0,1)$, it follows that
\begin{align}\label{equ9}
&x(t+1)=\begin{pmatrix}x_{1}(t+1) \\\vdots \\x_{n}(t+1)\end{pmatrix}\nonumber \\
&=\begin{pmatrix}fx_{avg}(t)+(1-f)(\alpha_{1}x_{1}(t)+ \sum\limits_{j=2}^{n} w_{1j}^{'}x_{j }(t)) \\\vdots \\fx_{avg}(t)+(1-f)(\alpha_{n}x_{n}(t)+ \sum\limits_{j=1}^{n-1} w_{nj}^{'}x_{j }(t))\end{pmatrix} \nonumber\\
&=\begin{pmatrix}\frac{f}{n}  {\textstyle \sum\limits _{k=1}^{n}} x_{k}(t) \\\vdots \\\frac{f}{n}  {\textstyle \sum\limits _{k=1}^{n}} x_{k}(t)\end{pmatrix}
+(1-f)\begin{pmatrix}\alpha_{1}x_{1}(t)+ \sum\limits_{j=2}^{n} w_{1j}^{'}x_{j }(t) \\\vdots \\\alpha_{n}x_{n}(t)+ \sum\limits_{j=1}^{n-1} w_{nj}^{'}x_{j }(t)\end{pmatrix}\nonumber\\
&=\begin{pmatrix}\frac{f}{n}  & \dots  & \frac{f}{n} \\\vdots& \ddots  &\vdots  \\\frac{f}{n}& \dots  &\frac{f}{n}
\end{pmatrix}\begin{pmatrix}x_{1}(t) \\\vdots \\x_{n}(t)\end{pmatrix}\nonumber\\
&+(1-f)\begin{pmatrix}\alpha _{1} &w_{12}^{'}  &\dots  &w_{1n}^{'} \\\vdots & \vdots  & \ddots  &\vdots  \\ \nonumber
w_{n1}^{'}& w_{n2}^{'} &\dots   &\alpha _{n}\end{pmatrix}\begin{pmatrix}x_{1}(t) \\\vdots \\x_{n}(t)\end{pmatrix}\\
&=\begin{pmatrix}(1-f)\alpha_{1}+\frac{f}{n}   & \dots  & (1-f)w_{1n}^{'}+\frac{f}{n} \\\vdots & \ddots   &\vdots  \\(1-f)w_{n1}^{'}+\frac{f}{n}   & \cdots  &(1-f)\alpha_{n}+\frac{f}{n}\end{pmatrix}
\begin{pmatrix}x_{1}(t) \\\vdots \\x_{n}(t)\end{pmatrix}.
 \end{align}
Let $$B=\begin{pmatrix}(1-f)\alpha_{1}+\frac{f}{n}   & \dots  & (1-f)w_{1n}^{'}+\frac{f}{n} \\\vdots & \ddots   &\vdots  \\(1-f)w_{n1}^{'}+\frac{f}{n}   & \cdots  &(1-f)\alpha_{n}+\frac{f}{n}\end{pmatrix},$$
$b_{ij}=(1-f)w_{ij}^{'}+\frac{f}{n}$  is the element at the $i$ th row and $j$ th column of the matrix $B$. Therefore,
\begin{align}
\sum_{j=1}^{n} b_{ij}&=(1-f)w_{i1}^{'}+\frac{f}{n} +...+(1-f)w_{in}^{'}+\frac{f}{n}\nonumber\\
&=(1-f)(w_{i1}^{'}+...+\alpha _{i}+...+w_{in}^{'})+\frac{f}{n} \cdot n\nonumber\\
&= 1-f+f\nonumber\\
&=1.\nonumber
\end{align}
As stated above,  $B$ is a row-stochastic matrix. Then Eq. (\ref{equ9}) is simply written as
\begin{equation}\label{equ10}
x(t+1)=Bx(t).
\end{equation}
Define $$Q\triangleq\begin{pmatrix}
 \frac{f}{n}  & \dots  & \frac{f}{n} \\\vdots& \ddots  &\vdots  \\\frac{f}{n}& \dots  &\frac{f}{n}
\end{pmatrix},$$
$$W^{'}\triangleq\begin{pmatrix}\alpha _{1} &w_{12}^{'}  &\dots  &w_{1n}^{'} \\\vdots & \vdots  & \ddots  &\vdots  \\
  w_{n1}^{'}& w_{n2}^{'} &\dots   &\alpha _{n}\end{pmatrix}\\.$$
  So we have $B=Q+(1-f)W^{'}$.
It is worth noting that  $Q$ and $W^{'}$  are fixed matrices, with $1-f$ being a constant in the interval $(0,1) $. Consequently, the matrix $ B $ is also a fixed matrix. The system (\ref{equ10}) can be regarded as the Degroot model, where  $B$ is equivalent to the weight matrix.

For any  $$b_{ij}=(1-f)w_{ij}^{'}+\frac{f}{n},$$
where $(1-f)\in (0,1)$, $w_{ij}^{'}\geq0$ and $\frac{f}{n}>0$, we have $b_{ij}>0$. Therefore, there must exist a $t^{*}$ such that the matrix power $B^{t^* }$ contains at least one strictly positive column.   Based on Lemma \ref{le1},  consensus can be reached in the system (\ref{equ10}).

Let
$\lambda=(\lambda_{1},\lambda_{2},...,\lambda_{n}),  {\textstyle \sum_{i=1}^{n}} \lambda_{i}=1,\lambda_{i}\geq0$, and $\lambda\,B=\lambda$.
\begin{align}
\lim_{t \to \infty} x(t)&=\lim_{t \to \infty} Bx(t-1)\nonumber\\
&=\lim_{t \to \infty}B^tx(0)\nonumber\\
&= \begin{pmatrix}\lambda_{1} & \dots  &\lambda_{n} \\\vdots  & \ddots  & \vdots \\
    \lambda_{1} & \cdots  &\lambda_{n} \end{pmatrix}\begin{pmatrix} x_{1}(0)\\\vdots  \\ x_{n}(0)\end{pmatrix}\nonumber\\
&=c\mathbf{1},\nonumber
\end{align}
where
$$c= {\textstyle \sum_{i=1}^{n}} \lambda_{i}x_{i}(0).$$

Therefore, the opinions in the model (\ref{equ4}) achieve consensus. The consensus value is $c= {\textstyle \sum_{i=1}^{n}} \lambda_{i}x_{i}(0).$
This completes the proof for Theorem \ref{th1}.
\end{proof}
\section{Simulation analysis and results}

In this section, simulations will be provided to verify the effectiveness of the proposed model. We will also conduct further investigation into the influence of group pressure and self-confidence on opinion evolution.

\begin{figure}[h]
\centering
\includegraphics[width=0.2\textwidth]{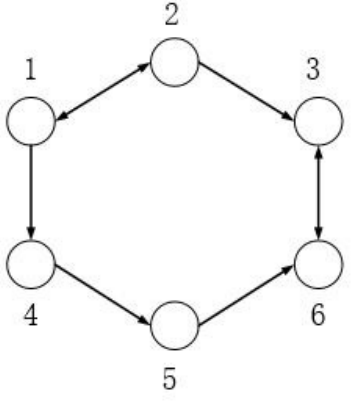}
\caption{A social network of six agents.}
\label{fig:3}
\end{figure}

Consider a simple social network of 6 agents.  The communication patterns among agents are depicted in Fig. \ref{fig:3}, and their trust relationships are detailed in Table \ref{table 1}. The self-confidence and original opinions of all agents are uniformly generated in the interval [0, 1], as listed below.$$\alpha=(0.32,0.63,0.1,0.84,0.76,0.55)^{T}.$$
$$x(0)=(0.14,0.8,0.4,0.9,0.2,0.36)^{T}.$$
\begin{table}[h]
\centering
\caption{Trust sets for each agent in Fig. \ref{fig:3}}
\label{table 1}
\scalebox{1.1}{
\begin{tabular}{cccc}\hline
			 &       $\mathcal{N}_{1}(i)$ &       $\mathcal{N}_{2}(i)$ &  \\   \hline
			1  &       $\{2,4\}$   &    $\{3,5\}$     \\	
			2  &       $\{1,3\}$     &   $\{4,6\}$      \\	
			3  &       $\{6\}$       &   $\emptyset$  \\	
			4  &        $\{5\}$ &  $\{6\}$  \\
			5  &        $\{6\}$     &   $\{3\}$  \\
			6  &        $\{3\}$             &  $\emptyset$  \\\hline
      \end{tabular}  }
\end{table}
We firstly set $f=0.5$, $r=0.8$.

According to Eq.  (\ref{equ6}), the weight matrix without considering two-step communication is as follows:
\begin{equation*}
W=\begin{pmatrix}0.32& 0.3386 & 0& 0.3414&0&0\\0.2101 &0.63&0.1599&0&0&0\\0&0&0.1&0&0&0.9\\0&0&0&0.84&0.16&0
\\0&0&0&0&0.76&0.24\\0&0&0.45&0&0&0.55
\end{pmatrix}.
\end{equation*}\\
According to Eq. (\ref{equ5}), the weight matrix  considering two-step communication is as follows:
\begin{equation*}
W^{'}=\begin{pmatrix}0.32& 0.2691 & 0.0764& 0.2713&0.0633&0\\0.1393 &0.63&0.106&0.0437&0&0.081\\0&0&0.1&0&0&0.9\\0&0&0&0.84&0.0963&0.0637
\\0&0&0.0726&0&0.76&0.1674\\0&0&0.45&0&0&0.55 \end{pmatrix}.
\end{equation*}\\
Note that,  agents assign the weight to their indirect neighbors, thereby trusting their opinions.
Hence, we have
\begin{equation*}
B=\begin{pmatrix}0.2433& 0.2179 & 0.1215& 0.219&0.115&0.0833\\0.153 &0.3983&0.1363&0.1052&0.0833&0.1239\\0.0833&0.0833&0.1333&0.0833&0.0833&0.5333\\0.0833&0.0833&0.08333&0.5033&0.1315&0.1152
\\0.0833&0.0833&0.1197&0.0833&0.4633&0.167\\0.0833&0.0833&0.3083&0.0833&0.0833&0.3583\end{pmatrix}.\\
\end{equation*}

\begin{figure}[h]
\centering
\includegraphics[width=0.5\textwidth]{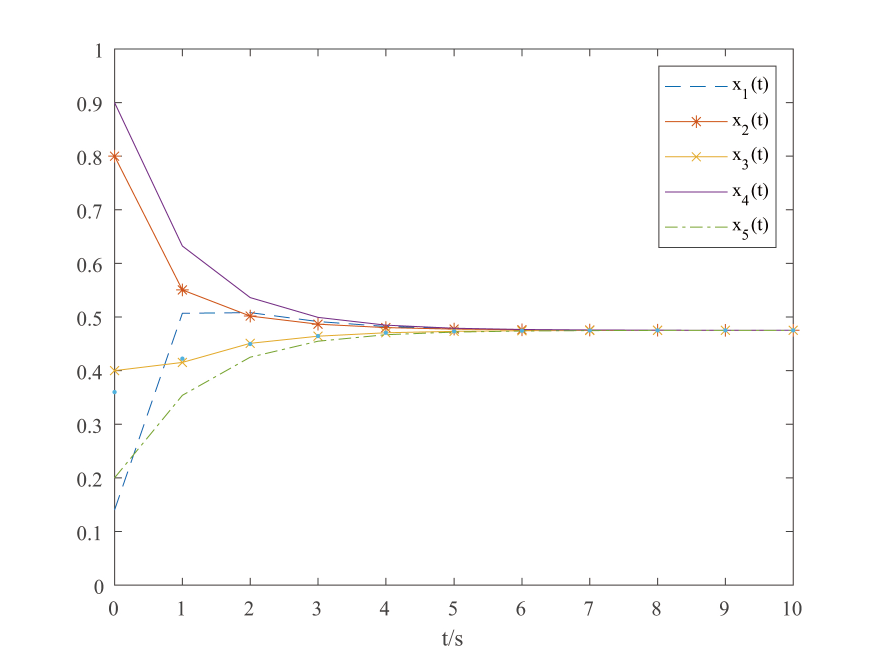}
\caption{The process of the opinions evolution in Fig. \ref{fig:3}.}
\label{fig:4}
\end{figure}

Based on Eq. (\ref{equ4}), the evolutionary opinion process  is illustrated in Fig. \ref{fig:4}. It can be demonstrated that the six agents ultimately reached a consensus.
\subsection{The impact of group pressure  on opinion dynamics}
In this subsection, it is explored how group pressure influences the process of opinion evolution. We will conduct two comparative simulation experiments to investigate the impact of group pressure, utilizing the same social network as depicted in Fig. \ref{fig:3}.

\textit{Experiment 1}:
In the first experiment, two scenarios are considered: one involves the influence of group pressure on the evolution of opinions, and the other scenario assumes group pressure is absent.  To ensure the isolation of other factors, all conditions in the experiment are kept the same except for the variable of group pressure. We set  group pressure $f=0.5$ and $f=0$, respectively.

\begin{figure}[ht]
\centering
\includegraphics[width=0.5\textwidth]{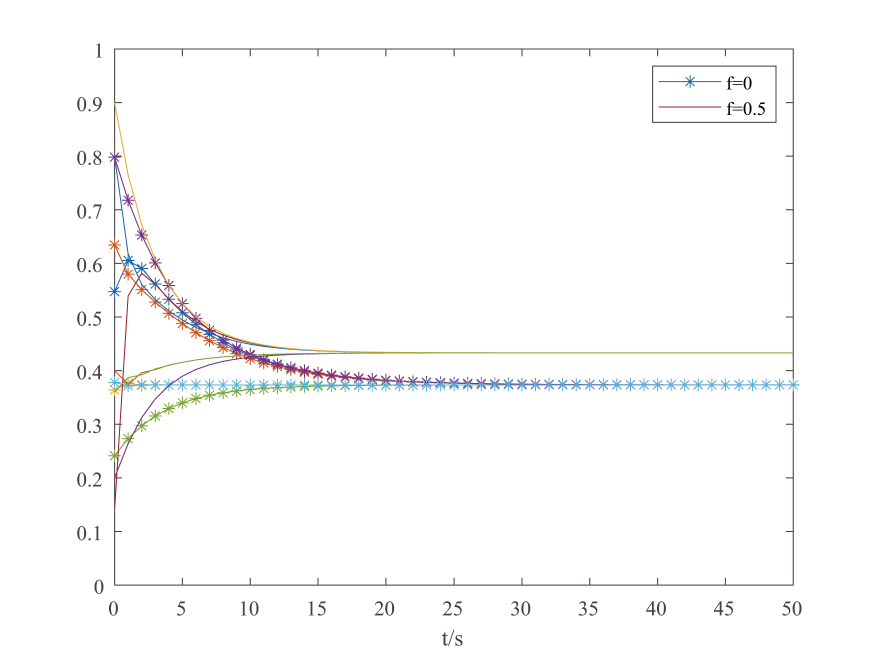}
\caption{The process of the opinions evolution with or without group pressure.}
\label{fig:5}
\end{figure}

The simulation result is shown in Fig. \ref{fig:5}. Notice that as for the convergence time of the group with or without group pressure, the former needs much less time than the latter. In other words, group pressure facilitates the group in reaching a consensus more easily.

\begin{figure}
    \centering
    \begin{subfigure}[a]{0.5\linewidth}
        \centering
        \includegraphics[width=\linewidth]{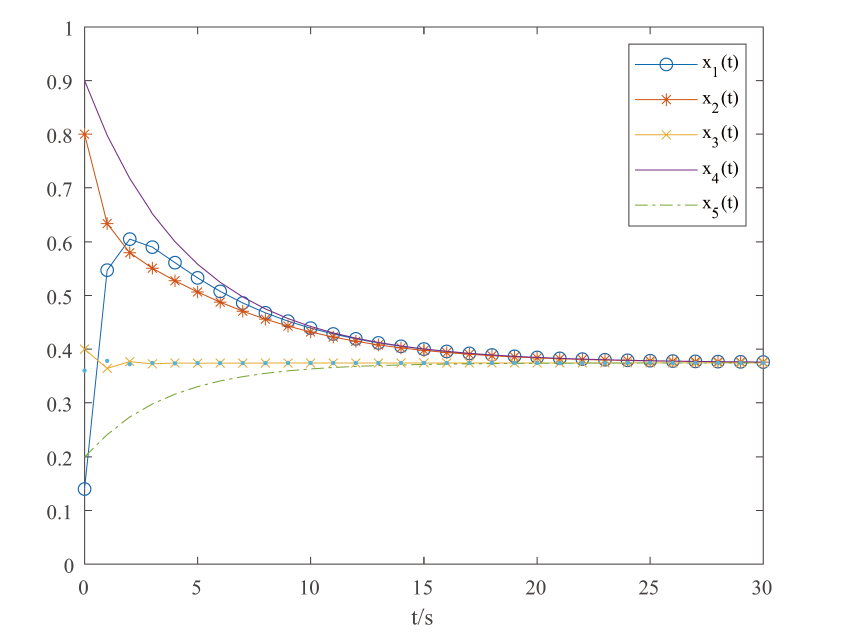}
		\caption{$f=0.001$}
		\label{subfigurea}
    \end{subfigure}%
    \begin{subfigure}[a]{0.5\linewidth}
        \centering
        \includegraphics[width=\linewidth]{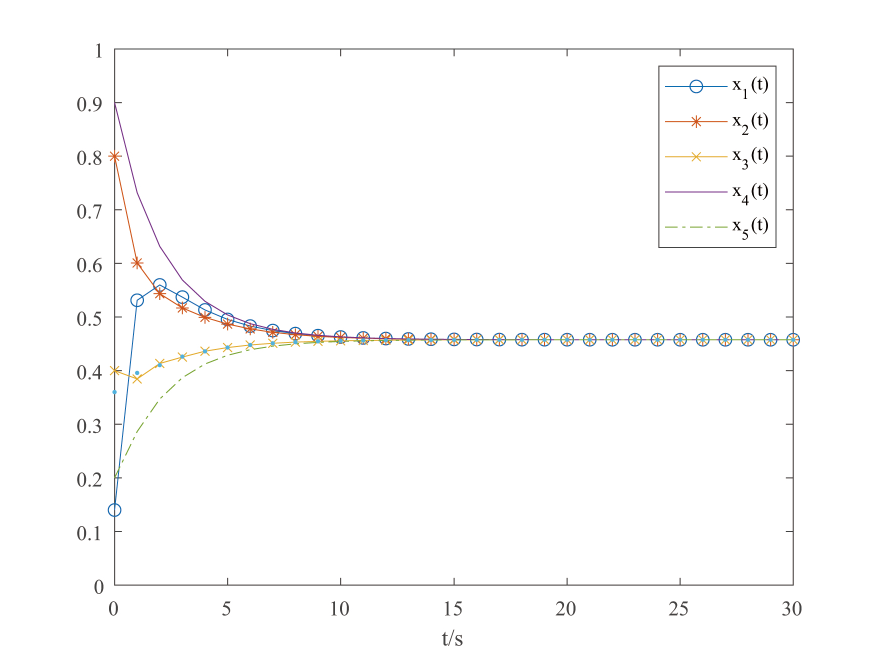}
		\caption{ $f=0.2$}
		\label{subfigureb}
    \end{subfigure}%

    \begin{subfigure}[a]{0.5\linewidth}
        \centering
        \includegraphics[width=\linewidth]{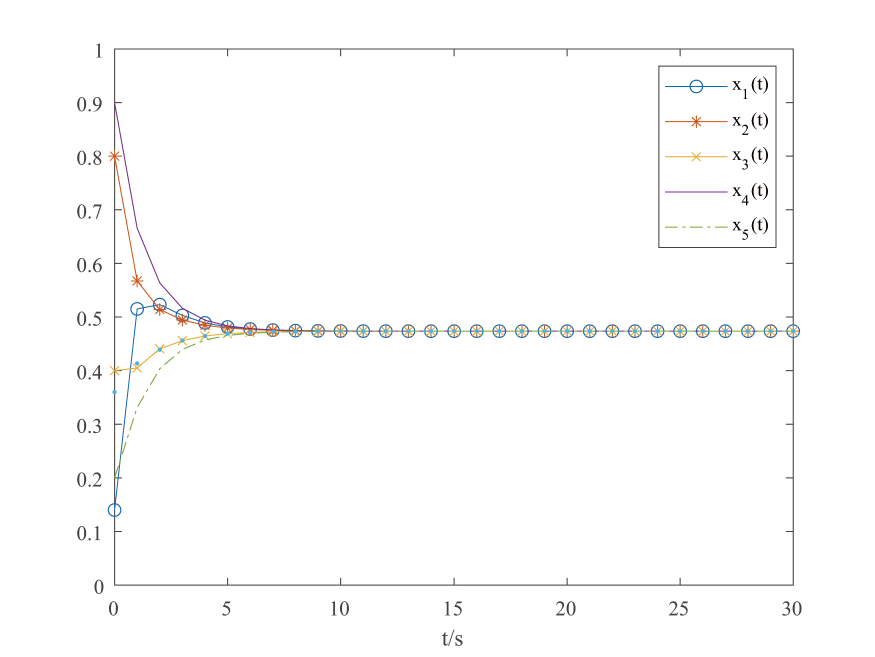}
		\caption{ $f=0.4$}
		\label{subfigurec}
    \end{subfigure}%
\begin{subfigure}[a]{0.5\linewidth}
        \centering
        \includegraphics[width=\linewidth]{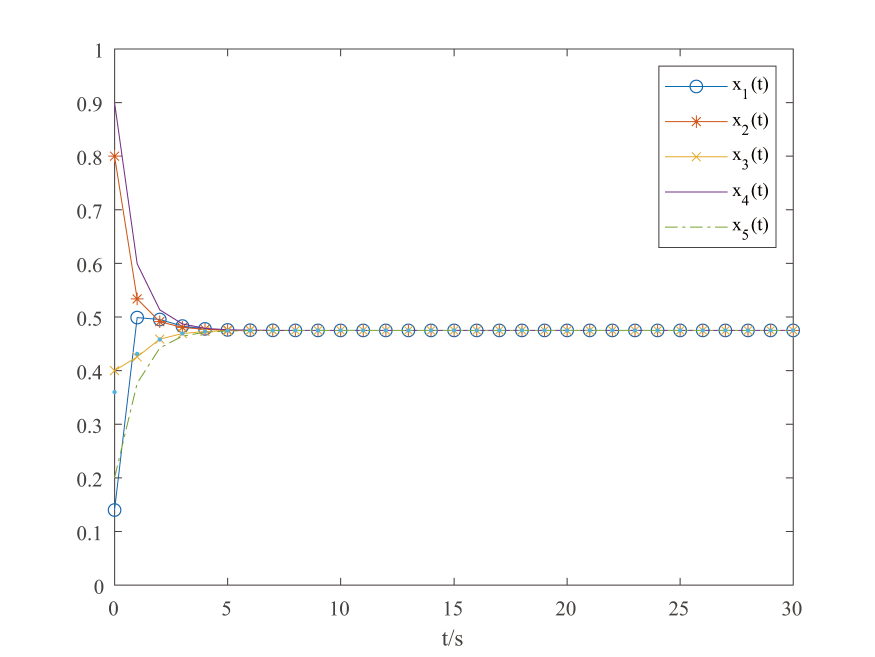}
		\caption{$f=0.6$}
		\label{subfigured}
    \end{subfigure}%

\begin{subfigure}[a]{0.5\linewidth}
        \centering
        \includegraphics[width=\linewidth]{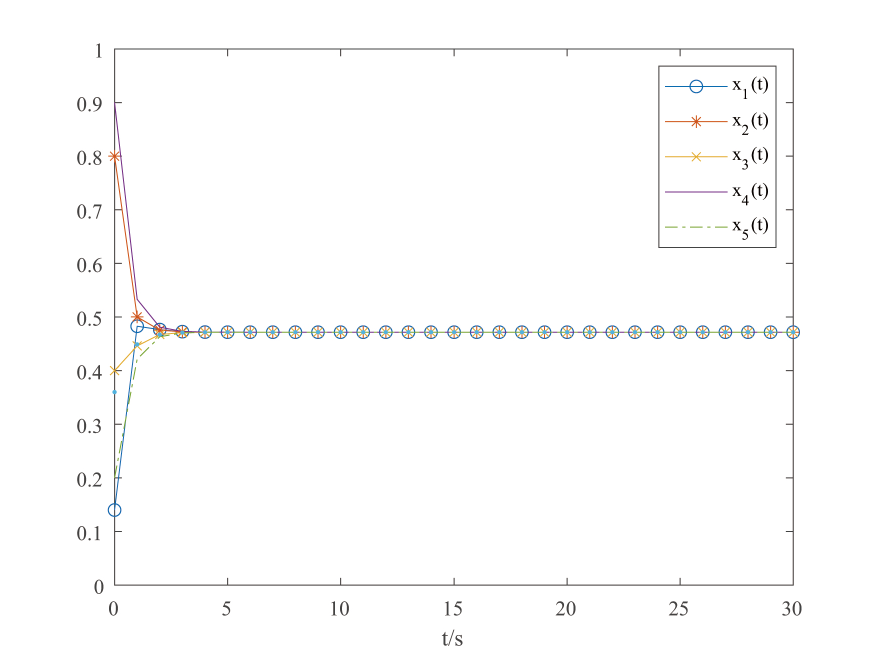}
		\caption{ $f=0.8$}
		\label{subfiguree}
    \end{subfigure}%
\begin{subfigure}[a]{0.5\linewidth}
        \centering
        \includegraphics[width=\linewidth]{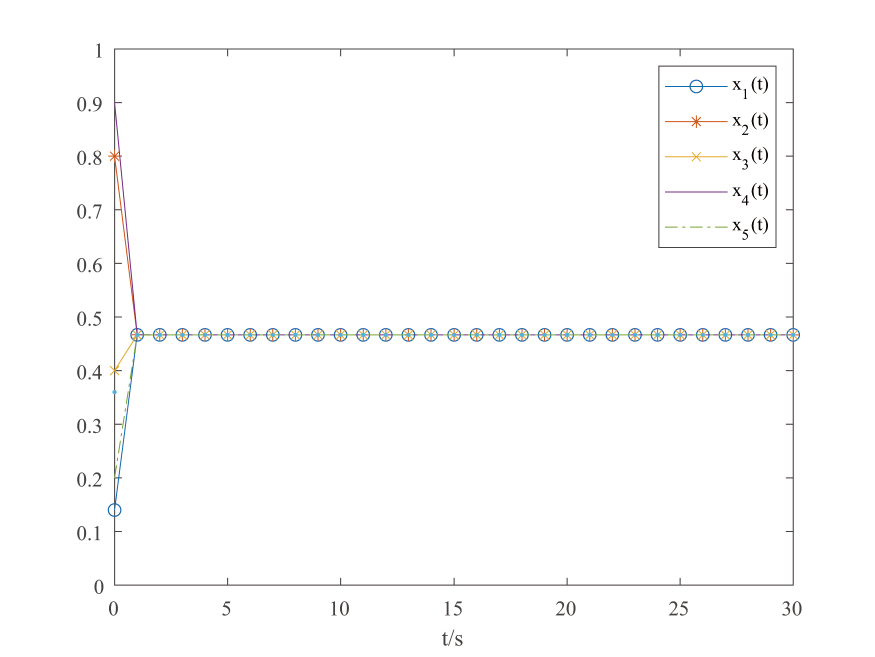}
		\caption{$f=0.999$}
		\label{subfiguref}
    \end{subfigure}%
	\caption{The process of the opinions evolution with different group pressure. (a) $f=0.001$; (b) $f=0.2$; (c) $f=0.4$; (d) $f=0.6$; (e) $f=0.8$; (f) $f=0.999$. }
	\label{figure6}
\end{figure}

\textit{Experiment 2}:
By further investigating the impact of group pressure, we compared the evolution of opinions under different group pressure level  with $f=0.001, 0.2, 0.4, 0.6, 0.8, 0.999$.
As shown in Fig. \ref{figure6}, the group finally reached a consensus in each of these cases.

It is obvious that the convergence time and group pressure are correlated. As the level of group pressure gradually increases, the time required for  agents in the proposed model to achieve consensus diminishes progressively. Interestingly, with very high group pressure level, consensus is rapidly attained  within the group. Therefore, it is
clear that the time for agents' opinions to achieve consensus decreases as group pressure increases. This further validates the results obtained in the first experiment.

\begin{figure}[htbp]
    \centering
    \begin{subfigure}[a]{0.5\linewidth}
        \centering
\includegraphics[width=\linewidth]{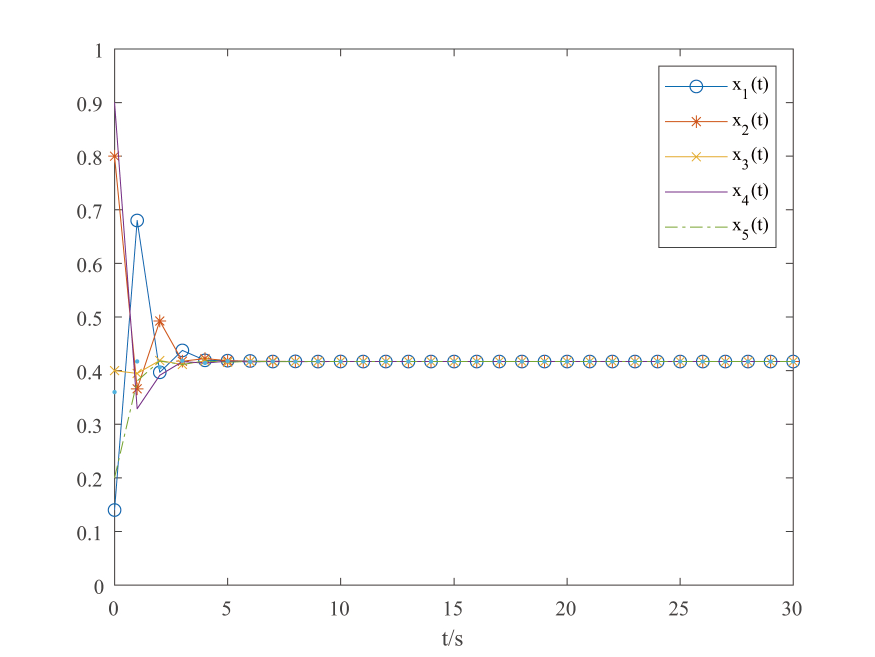}	\caption{$\alpha_{i}=0.1$}
		\label{sub7a}
    \end{subfigure}%
    \begin{subfigure}[a]{0.5\linewidth}
        \centering  \includegraphics[width=\linewidth]{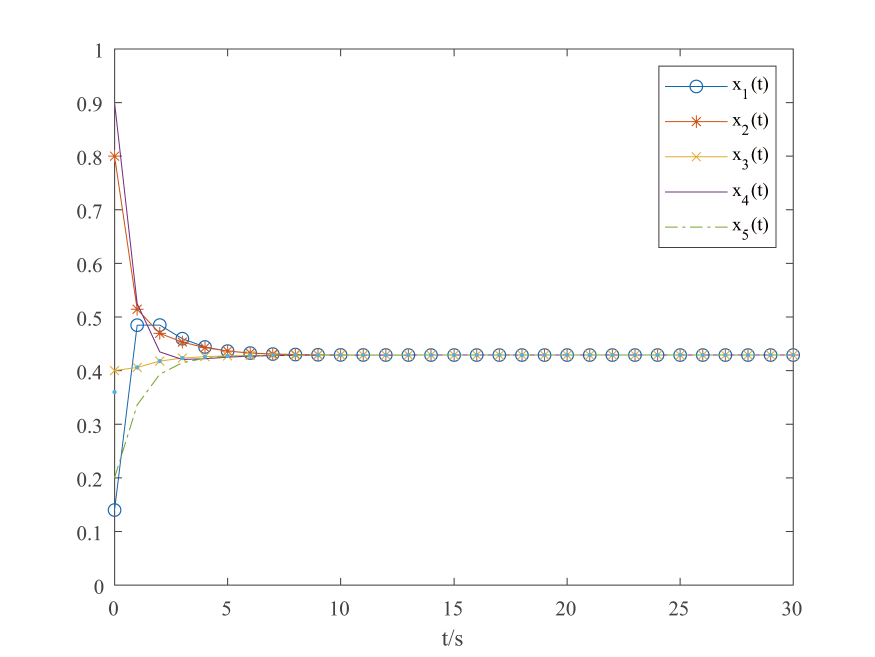}
		\caption{ $\alpha_{i}=0.3$}
		\label{sub7b}
    \end{subfigure}%

    \begin{subfigure}[a]{0.5\linewidth}
        \centering \includegraphics[width=\linewidth]{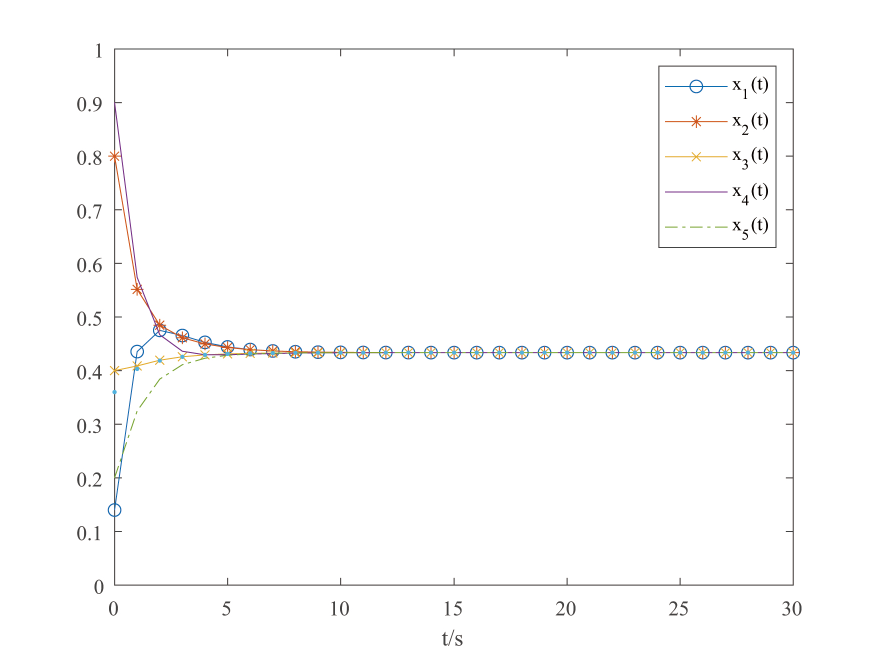}
		\caption{ $\alpha_{i}=0.6$}
		\label{sub7c}
    \end{subfigure}%
\begin{subfigure}[a]{0.5\linewidth}
        \centering \includegraphics[width=\linewidth]{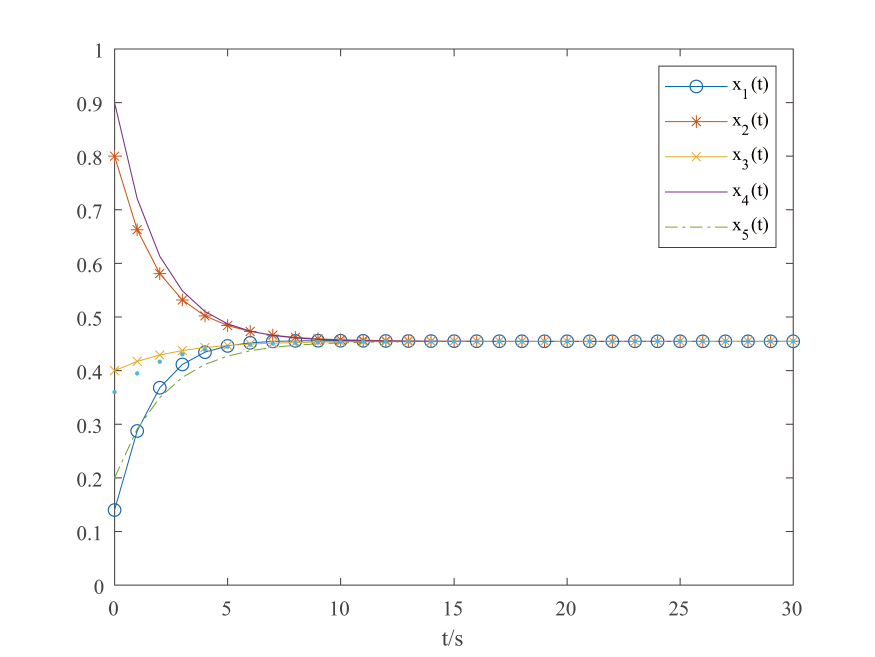}
		\caption{ $\alpha_{i}=0.9$}
		\label{sub7d}
    \end{subfigure}%
    \caption{The process of the opinions evolution with different self-confidence. (a) $\alpha_{i}=0.1$; (b) $\alpha_{i}=0.3$; (c) $\alpha_{i}=0.6$; (d) $\alpha_{i}=0.9$. }
	\label{figure7}
\end{figure}
Based on the two simulation experiments,  group pressure facilitates consensus formation within social networks, and the level of group pressure is inversely proportional to the time required for  group opinions to reach consensus.
\subsection{The impact of self-confidence on opinion dynamics}

Self-confidence is a key factor in the evolution of opinions and affects the speed at which consensus is reached. In this subsection, we discuss the influence of  self-confidence on opinion dynamics.

To rule out the impact of original opinions, we use the same initial profile as in the preceding subsection.
For convenience, it is assumed that all agents have the same level of self-confidence and investigate the evolution of opinions under three different cases: namely, open-minded $\alpha_{i} = 0.1$, moderate-minded $\alpha_{i} = 0.5$,
 and closed-minded $\alpha_{i} =0. 9$. Set the group pressure as  $f = 0.3$.
As depicted in Fig. \ref{figure7}, the time taken to reach consensus gradually increases with  the higher level of self-confidence. It is obvious that the higher  of self-confidence, the more difficult it is for opinions to reach consensus. The simulation results show that  self-confidence effectively inhibits the formation of consensus.

\section{Conclusion}

In this paper,
combining self-confidence, group pressure and two-step communication, a novel two-step communication opinion dynamics model is proposed based on the DeGroot model.
In the proposed model, agents are influenced by group pressure when updating their opinions and their neighbors are classified into direct neighbors and indirect neighbors for two-step communication.
This overcomes the limitations of the DeGroot model, in which individuals communicate only with their own neighbors, thereby ignoring the neighbors of their neighbors and public opinion.
Then, we explore the impact of group pressure and individual self-confidence on opinion consensus in a simple social network. The simulation results demonstrate that group pressure  promotes the formation of consensus, while  self-confidence suppresses it. These findings provide valuable insights for researchers assessing the practical application of group pressure and self-confidence. Further work will focus on the theoretical analysis of the effects of time-varying network topologies and the construction of opinion dynamics models in uncertain environments.

\ifCLASSOPTIONcaptionsoff
  \newpage
\fi



%
\bibliographystyle{IEEEtran}
\bibliography{ref}

\begin{thebibliography}{10}
\providecommand{\url}[1]{#1}
\csname url@samestyle\endcsname
\providecommand{\newblock}{\relax}
\providecommand{\bibinfo}[2]{#2}
\providecommand{\BIBentrySTDinterwordspacing}{\spaceskip=0pt\relax}
\providecommand{\BIBentryALTinterwordstretchfactor}{4}
\providecommand{\BIBentryALTinterwordspacing}{\spaceskip=\fontdimen2\font plus
\BIBentryALTinterwordstretchfactor\fontdimen3\font minus
  \fontdimen4\font\relax}
\providecommand{\BIBforeignlanguage}[2]{{%
\expandafter\ifx\csname l@#1\endcsname\relax
\typeout{** WARNING: IEEEtran.bst: No hyphenation pattern has been}%
\typeout{** loaded for the language `#1'. Using the pattern for}%
\typeout{** the default language instead.}%
\else
\language=\csname l@#1\endcsname
\fi
#2}}
\providecommand{\BIBdecl}{\relax}
\BIBdecl

\bibitem{TIAN2018213}
Y.~Tian and L.~Wang, ``Opinion dynamics in social networks with stubborn
  agents: An issue-based perspective,'' \emph{Automatica}, vol.~96, pp.
  213--223, Oct. 2018.

\bibitem{wulixue}
C.~Castellano, S.~Fortunato, and V.~Loreto, ``Statistical physics of social
  dynamics,'' \emph{Rev. Mod. Phys.}, vol.~81, no.~2, May 2009.

\bibitem{dandekar2013biased}
P.~Dandekar, A.~Goel, and D.~T. Lee, ``Biased assimilation, homophily, and the
  dynamics of polarization,'' \emph{Proc. Natl. Acad. Sci.}, vol. 110, no.~15,
  pp. 5791--5796, Mar. 2013.

\bibitem{10049708}
Q.~Liu and L.~Chai, ``The memory influence on opinion dynamics in coopetitive
  social networks: Analysis, application, and simulation,'' \emph{IEEE Trans.
  Control Netw. Syst}, vol.~10, no.~4, pp. 1867--1878, Feb. 2023.

\bibitem{li2019clustering}
X.~Li, J.~Zhang, Y.~Zou, and S.~Guan, ``Clustering and bellerophon state in
  kuramoto model with second-order coupling,'' \emph{Chaos}, vol.~29, no.~4,
  Apr. 2019.

\bibitem{citation-key}
J.~L. Moreno, ``Who shall survive? a new approach to the problem of human
  interrelations,'' Nervous and Mental Disease, Washington, D.C., 1934.

\bibitem{wiener2019cybernetics}
N.~Wiener, ``Cybernetics,'' Hermann, Paris, 1948.

\bibitem{geyer1995challenge}
F.~Geyer, ``The challenge of sociocybernetics,'' \emph{Kybernetes}, vol.~24,
  no.~4, pp. 6--32, Jun. 1995.

\bibitem{zheng1992current}
D.~Zheng and Y.~Zheng, ``The current state and developing trends of deds
  theory,'' \emph{Acta Automatica Sinica}, vol.~2, no.~18, pp. 129--142, Apr.
  1992.

\bibitem{jin2023distributed}
N.~Jin, J.~Xu, and H.~Zhang, ``Distributed optimal consensus control of
  multi-agent systems involving state and control dependent multiplicative
  noise,'' \emph{IEEE Trans. Autom. Control}, vol.~68, no.~12, pp. 7787--7794,
  Feb. 2023.

\bibitem{qin2021multiagent}
J.~Qin, Q.~Ma, P.~Yi, and L.~Wang, ``Multiagent interval consensus with
  flocking dynamics,'' \emph{IEEE Trans. Autom. Control}, vol.~67, no.~8, pp.
  3965--3980, Oct. 2021.

\bibitem{multi2021}
L.~Shi, Y.~Cheng, J.~Shao, and X.~Zhang, ``Collective behavior of multileader
  multiagent systems with random interactions over signed digraphs,''
  \emph{IEEE Trans. Control Netw. Syst}, vol.~8, no.~3, pp. 1394--1405, Mar.
  2021.

\bibitem{valcher2017consensus}
M.~E. Valcher and I.~Zorzan, ``On the consensus of homogeneous multi-agent
  systems with positivity constraints,'' \emph{IEEE Trans. Autom. Control},
  vol.~62, no.~10, pp. 5096--5110, Apr. 2017.

\bibitem{liu2023event}
M.~Li, Z.~Wu, F.~Deng, and B.~Guo, ``Active disturbance rejection control to
  consensus of second-order stochastic multiagent systems,'' \emph{IEEE Trans.
  Control Netw. Syst}, vol.~10, no.~2, pp. 993--1004, Jun. 2023.

\bibitem{degroot1974reaching}
M.~H. DeGroot, ``Reaching a consensus,'' \emph{J. Am. Stat. Assoc.}, vol.~69,
  no. 345, pp. 118--121, Oct. 1974.

\bibitem{friedkin1990social}
N.~E. Friedkin and E.~C. Johnsen, ``Social influence and opinions,'' \emph{J.
  Math. Sociol.}, vol.~15, no. 3-4, pp. 193--206, 1990.

\bibitem{deffuant2000mixing}
G.~Deffuant, D.~Neau, F.~Amblard, and G.~Weisbuch, ``Mixing beliefs among
  interacting agents,'' \emph{Adv. Complex Syst.}, vol.~3, pp. 87--98, 2000.

\bibitem{hegselmann2002opinion}
R.~Hegselmann and U.~Krause, ``Opinion dynamics and bounded confidence models,
  analysis, and simulation,'' \emph{J. Artif. Soc. Soc. Simul.}, vol.~5, no.~3,
  2002.

\bibitem{wanglong}
L.~Wang, Y.~Tian, and J.~Du, ``Opinion dynamics in social networks,''
  \emph{Scientia Sinica Informationis}, vol.~48, no.~1, pp. 3--23, Jan. 2018.

\bibitem{bernardo2024bounded}
C.~Bernardo, C.~Altafini, A.~Proskurnikov, and F.~Vasca, ``Bounded confidence
  opinion dynamics: A survey,'' \emph{Automatica}, vol. 159, p. 111302, Oct.
  2024.

\bibitem{bipartite2021}
Y.~Zou and K.~Xia, ``Bipartite consensus of opinion dynamics through delivering
  credible information,'' \emph{IEEE Trans. Control Netw. Syst}, vol.~8, no.~2,
  pp. 781--790, Nov. 2021.

\bibitem{expressed}
W.~Xia, H.~Liang, and M.~Ye, ``Asynchronous expressed and private opinion
  dynamics on influence networks,'' \emph{IEEE Trans. Control Netw. Syst},
  vol.~10, no.~2, pp. 544--555, Jun. 2023.

\bibitem{xing2021novel}
Y.~Xing, Y.~Hong, and H.~Fang, ``A novel opinion model for complex
  macro-behaviors of mass opinion,'' \emph{Sci. China Inf. Sci.}, vol.~64,
  no.~2, p. 129205, Feb. 2021.

\bibitem{su2017noise}
W.~Su, G.~Chen, and Y.~Hong, ``Noise leads to quasi-consensus of
  hegselmann--krause opinion dynamics,'' \emph{Automatica}, vol.~85, pp.
  448--454, Nov. 2017.

\bibitem{he2021discrete}
G.~He, J.~Liu, H.~Hu, and J.~Fang, ``Discrete-time signed bounded confidence
  model for opinion dynamics,'' \emph{Neurocomputing}, vol. 425, pp. 53--61,
  Dec. 2021.

\bibitem{anderson2019recent}
B.~D. Anderson and M.~Ye, ``Recent advances in the modelling and analysis of
  opinion dynamics on influence networks,'' \emph{Int. J. Autom. Comput.},
  vol.~16, no.~2, pp. 129--149, Feb. 2019.

\bibitem{ye2019influence}
M.~Ye, Y.~Qin, A.~Govaert, B.~D. Anderson, and M.~Cao, ``An influence network
  model to study discrepancies in expressed and private opinions,''
  \emph{Automatica}, vol. 107, pp. 371--381, Sep. 2019.

\bibitem{javarone2014social}
M.~A. Javarone, ``Social influences in opinion dynamics: the role of
  conformity,'' \emph{Physica A}, vol. 414, pp. 19--30, Nov. 2014.

\bibitem{Cheng2019}
C.~Cheng and C.~Yu, ``Opinion dynamics with bounded confidence and group
  pressure,'' \emph{Physica A}, vol. 532, Oct. 2019.

\bibitem{Zhang2024}
S.~Zhang, B.~Liu, and L.~Chai, ``Analysis of modified hegselmann-krause opinion
  dynamics based on conformity,'' \emph{Control and Decision}, vol.~39, pp.
  965--974, Mar 2024.

\bibitem{qian2011adaptive}
C.~Qian, J.~Cao, J.~Lu, and J.~Kurths, ``Adaptive bridge control strategy for
  opinion evolution on social networks,'' \emph{Chaos}, vol.~21, no.~2, Jun.
  2011.

\bibitem{connected}
N.~A. Christakis and J.~H. Fowler, ``Connected: The surprising power of our
  social networks and how they shape our lives,'' New York, 2009.

\bibitem{atwostep2020}
Q.~Zhou, Z.~Wu, A.~H. Altalhi, and F.~Herrera, ``A two-step communication
  opinion dynamics model with self-persistence and influence index for social
  networks based on the degroot model,'' \emph{Inf. Sci.}, vol. 519, pp.
  363--381, May 2020.

\bibitem{DONG2022218}
Q.~Dong, Q.~Sheng, L.~Martínez, and Z.~Zhang, ``An adaptive group decision
  making framework: Individual and local world opinion based opinion
  dynamics,'' \emph{Inf. Fusion}, vol.~78, pp. 218--231, Sep. 2022.

\bibitem{dong2017managing}
Y.~Dong, Z.~Ding, L.~Mart{\'\i}nez, and F.~Herrera, ``Managing consensus based
  on leadership in opinion dynamics,'' \emph{Inf. Sci.}, vol. 397, pp.
  187--205, Aug. 2017.

\end{thebibliography}
\end{document}